\documentclass[superscriptaddress,11pt]{article}%
\usepackage{authblk}
\usepackage{array}
\usepackage{type1cm}
\usepackage{lettrine}
\usepackage{graphicx}
\usepackage{geometry}
\usepackage{psfrag}
\usepackage{amsmath}
\usepackage{amsfonts}
\usepackage{appendix}
\usepackage[margin=10pt,font=footnotesize,labelfont=sc,format=hang,labelformat=parens]%
{caption}
\usepackage{amssymb}%
\providecommand{\U}[1]{\protect\rule{.1in}{.1in}}

\numberwithin{equation}{section}
\begin{document}

\title{The generator of spatial diffeomorphisms in 
the Koslowski- Sahlmann representation}
\author{Madhavan Varadarajan}
\affil{Raman Research Institute\\Bangalore-560 080, India}
\maketitle

\begin{abstract}

A generalization of the  representation, underlying the discrete spatial geometry of  Loop Quantum Gravity, 
to accomodate states labelled by smooth spatial 
geometries,  was discovered by  Koslowski and further studied by Sahlmann. We show how to construct 
the diffeomorphism constraint operator in this Koslowski- Sahlmann (KS) representation from 
suitable  connection and triad dependent operators.
We show that  the KS representation supports the action of hitherto unnoticed
``background exponential'' operators which, in contrast to the holonomy- flux operators, change the 
smooth spatial geometry label
of the states. 
These operators are shown to be quantizations of certain connection dependent functions
 and play a key role in the construction of the diffeomorphism constraint operator.

\end{abstract}

\thispagestyle{empty}
\let\oldthefootnote\thefootnote\renewcommand{\thefootnote}{\fnsymbol{footnote}}
\footnotetext{Email: madhavan@rri.res.in}
\let\thefootnote\oldthefootnote

\section{Introduction}

Loop Quantum Gravity is based on the canonical quantization of a classical Hamiltonian description of
gravity in terms of an $SU(2)$ connection and its conjugate electric (triad) field. The basic operators of the theory
are holonomies of the connection around loops in the Cauchy slice $\Sigma$  and electric fluxes through surfaces therein. 
The LQG representation yields a discrete spatial geometry 
at the quantum level
as exemplified by the discrete spectra of triad dependent operators such as the volume and the area
\cite{discrete}.

In order to capture the effective smoothness of classical spatial geometries, Koslowski \cite{kos}
constructed a representation of the holonomy- flux algebra in which the flux operator action
is augmented by a `classical flux'  contribution and the holonomy operator action is unchanged.
This representation was further studied with a view towards imposing the $SU(2)$ Gauss Law and the 
diffeomorphism constraints of gravity by Sahlmann \cite{hanno}. In this Koslowski- Sahlmann (KS) representation,
states acquire an additional label corresponding to a smooth `background' triad field, the additional contribution
to the flux operator coming precisely from the flux of this background electric field.
As seen in equations (\ref{hks})- (\ref{eks}) below, neither the holonomy nor the flux operators change this
background field. This feature gives rise to the following conundrum.

Finite spatial diffeomorphisms are unitarily represented in the KS representation \cite{hanno} and their action
on a state labelled by some background triad field yields a state labelled with a new background field, the new
label being the image of the old one under the spatial diffeomorphism. On the other hand, the recent construction 
of the diffeomorphism constraint in LQG \cite{mealok} from the basic holonomy- flux operators
of LQG in such a way that it generates diffeomorphisms, suggests that a similar construction should 
exist in the KS representation. Since the diffeomorphism constraint depends on the connection and electric field,
and since the holonomy-flux variables seperate points on the phase space of connections and electric fields,
one may expect that, similar to Reference \cite{mealok}, the diffeomorphism constraint operator
should be expressed in terms of (a limit of) holonomy- flux operators. However, this is at odds with the feature mentioned
in the previous paragraph because diffeomorphisms do change the background triad field whereas the holonomy and flux
operators do not.

In this work we resolve this conundrum by identifying a set of connection dependent quantities
which, upon quantization, yield operators which do change the background triad. Using these operators
together with the holonomy- flux operators yields a satisfactory construction of the diffeomorphism
constraint. In section 2, we review the KS representation, identify the new ``background exponential''
functions and construct their quantization. In section 3, we indicate how to construct the 
diffeomorphism constraint from the enlarged holonomy-flux- background exponential algebra.
Section 4 contains our concluding remarks and a description of results recently obtained
such as   the  crucial role of background exponentials in an improved treatement of the imposition of the
Gauss Law and diffeomorphism constraints \cite{miguelme} and the construction of the KS representation for two
dimensional Parameterized
Field Theory in such a way as to support its dynamics
\cite{sandipanc}.

In what follows we assume familiarity with standard LQG, and, in section 3, with the results of Reference \cite{mealok}.

\section{Background exponentials}

The classical phase space variables for LQG are the connection $A_a^i$, $i$ being an internal $SU(2)$ Lie algebra valued 
index, and the conjugate 
densitized  electric (triad)  field $E^a_i$ with Poisson brackets 
$\{A_a^i(x),E^b_j (y)\}=  8\pi \gamma G \delta (x,y)\delta^i_j$ where $\gamma$ is the Barbero- Immirzi parameter.
We shall choose units such that $8\pi \gamma G= \hbar= c=1$.

The kinematical Hilbert space of standard LQG is spanned by the orthormal basis of spin network states $\{|s\rangle\}$.
Let the dense domain of the finite linear span of spinnets be ${\cal D}$. Let $\hat O$ be  an operator 
from ${\cal D}$ to ${\cal D}$ so that  $\hat O|s\rangle$ is a finite linear combination of spinnets i.e.  
$\hat O|s\rangle = \sum_IO^{(s)}_I|s_I\rangle$ where $O_I$ are the complex  coefficients in the sum over the spinnets
$|s_I\rangle$. 
It is useful to introduce the notation $|{\hat O}s\rangle$ to denote this linear combination of spinnets so that 
we have that 
\begin{equation} 
|{\hat O}s\rangle:= \hat O|s\rangle = \sum_IO^{(s)}_I|s_I\rangle .
\label{o}
\end{equation}

The Koslowski- Sahlmann kinematic Hilbert space is then spanned by states  which have, in addition to their LQG
spinnet label, an additional label ${\bf E}^a_{i}$ where ${\bf E}^a_{i}$ is a smooth ``background'' electric field.
We denote such a state by $|s,{\bf E}\rangle$. These 
states for all $s, {\bf E}^a_{i}$ provide an orthonormal basis for the Koslowski- Sahlmann kinematic Hilbert space
so that the inner product between two such KS spinnets in this Hilbert space is
\begin{equation}
\langle s^{\prime},{\bf E}^{\prime }_{}|s,{\bf E}\rangle = <s|s^{\prime}> 
\delta_{{\bf E}^{\prime a}_{i}, {\bf E}^a_{i}},
\end{equation}
where $ <s|s^{\prime}>$ is just the standard LQG inner product and the second factor is the Kronecker delta which 
vanishes unless the two background fields agree in which case it equals unity.

The holonomy- flux operators act on the KS spinnets as:
\begin{eqnarray}
{\hat h}^A_{\alpha\;B}|s,{\bf E}_{}\rangle &=& \vert{\hat h}^A_{\alpha\;B}s,{\bf E}_{}\rangle ,
\label{hks}
\\
{\hat E}_S(f)|s,{\bf E}_{}\rangle &=& |{\hat E}_S(f)s,{\bf E}_{}\rangle + {\bf E}_{S}(f)|s,{\bf E}_{}\rangle .
\label{eks}
\end{eqnarray}
Here, our notation is as follows.
${\hat h}^A_{\alpha\;B}$
is the $A,B$ component of the holonomy operator around the loop $\alpha$ in the $j=\frac{1}{2}$ representation.  
${\hat E}_S(f)$ is  the electric flux operator obtained by integrating the electric field
with respect to the Lie algebra valued function $f^i$ over the surface $S$ i.e.
\begin{equation}
{\hat E}_S(f)= \int_S f^i{\hat E}^a_i \eta_{abc} ,
\label{fluxdef}
\end{equation}
where $\eta_{abc}$ is the Levi- Civita tensor of weight $-1$.
 We have used an obvious generalization of the notation of equation (\ref{o}) wherein 
given an operator ${\hat O}$ with action 
$\hat O|s\rangle = \sum_IO^{(s)}_I|s_I\rangle$  in standard LQG, we have defined the state $|{\hat O}s, {\bf E}\rangle$
in the KS representation through
\begin{equation} 
|{\hat O}s ,{\bf E}_{}\rangle:= \sum_IO^{(s)}_I|s_I,{\bf E}_{} \rangle .
\label{oks}
\end{equation}
Similar to equation (\ref{fluxdef}), in equation (\ref{eks}), 
${\bf E}_{S}(f)$ denotes the 
flux of ${\bf E}_{}^i$ smeared with $f^i$ through the surface $S$ i.e.
\begin{equation}
{\bf E}_S(f)= \int_S f^i{\bf E}^a_i \eta_{abc}.
\label{bfluxdef}
\end{equation}
Thus, 
the action of the electric flux operator in the KS representation obtains, 
in addition to the standard LQG like first term, an extra background electric flux contribution.
We note here that neither the holonomy nor the flux operators change the background electric field label of the state.

Next, we define an operator ${\hat H}_{\bf E}$ which changes the background electric field by translating it 
through the amount ${\bf E}^a_i$ so that: 
\begin{equation}
{\hat H}_{{\bf E}} |s, {\bf E}_0\rangle = |s, {\bf E}+{\bf E}_0\rangle .
\label{defheks}
\end{equation}
It is easily verified that this operator commutes with the holonomy operators, that its commutator
with the flux operator is 
\begin{equation}
[{\hat H}_{{\bf E}}, {\hat {E}}_S(f) ]= -{\bf E}_S(f) {\hat H}_{{\bf E}} ,
\label{beflux}
\end{equation}
and that two such operators ${\hat H}_{{\bf E}_1},{\hat H}_{{\bf E}_2}$ commute with each other.

Next, we note that these  operators can be constructed as quantizations of certain classical functions
which we call `background exponentials' for reasons which are obvious from the considerations below.
Each such  function depends only on the connection, is  labelled by some background electric field 
${\bf E}^a_i$ and is denoted by ${H}_{\bf E}$.
We define ${H}_{\bf E}$ as:
\begin{equation}
H_{{\bf E}}(A)= \exp (i\int_{\Sigma}A^i_a{\bf E}_i^a) .
\label{defbe}
\end{equation}
Clearly, these functions Poisson commute with each other as well as with the holonomies. 
The only additional non- trivial Poisson bracket is
\begin{equation}
\{H_{{\bf E}}(A), E_S(f)\}= i {\bf E}_S(f) H_{{\bf E}}(A) .
\label{bepb}
\end{equation}
It follows that the operator action (\ref{defheks}) provides a representation of the Poisson bracket
(\ref{bepb}) through  equation (\ref{beflux}). 

In summary: the operator ${\hat H}_{{\bf E}}$ defined in equation (\ref{defheks}) is the quantum
correspondent of the classical phase space ``background exponential''
function $H_{{\bf E}}(A)$ defined in equation (\ref{defbe}).

\section{The diffeomorphism constraint}

As shown in \cite{hanno}, in order for the holonomy- flux algebra to transform correctly under the unitary 
action of finite spatial diffeomorphisms, it is essential that the background electric field label of any KS state
transforms to its diffeomorphic image under this action. More precisely, as shown in \cite{hanno},
the unitary operator ${\hat U}(\phi)$ corresponding to the finite diffeomorphism $\phi$ acts on KS spinnets 
through:
\begin{equation}
{\hat U}(\phi )|s, {\bf E}\rangle 
= |{\hat U}(\phi ) s, \phi^*{\bf E}\rangle
= |\phi\circ s, \phi^*{\bf E}\rangle   ,
\label{uphiks}
\end{equation}
where $\phi\circ s$ is just the standard LQG diffeomorphic image of the spinnet $s$ and 
$\phi^*{\bf E}$ is the push forward of the background electric field under $\phi$.

In the classical theory, diffeomorphisms along the vector field $N^a$ 
are generated 
by the 
diffeomorphism constraint  $D({\vec N})$ where:
\begin{equation}
D({\vec N})= 
\int_{\Sigma}( {\cal L}_{\vec N} A_b^i )\;  E^b_i .
\label{defdn}
\end{equation}
 In \cite{mealok}, this operator was constructed
in standard LQG  out of the standard holonomy- flux operators by first constructing finite triangulation 
approximants which are well defined on the kinematic Hilbert space and then defing their
continuum limit action on the Lewandowski- Marolf (LM) habitat \cite{donjurek}.
In the KS representation, a construction involving only the holonomy- flux operators cannot generate
diffeomorphisms of the background electric field label because, from equations (\ref{hks})- (\ref{eks}),
these operators do not change the  background electric field. We now show that finite triangulation 
approximants to the diffeomorphism constraint which involve the background exponential operators
can be constructed in such a way as to correctly generate the action of diffeomorphisms in the KS representation.

We start by rewriting $D({\vec N})$ as
\begin{eqnarray}
D({\vec N}) &=& \int_{\Sigma}({\cal L}_{\vec N}A_b^i)\;E^b_i \nonumber\\
&=& \int_{\Sigma}({\cal L}_{\vec N}A_b^i)\;(E^b_i- {\bf E}_i^b) 
+\int_{\Sigma}({\cal L}_{\vec N}A_b^i)\;{\bf E}^b_i \\
&:=&  D^{\bf E}({\vec N}) +\int_{\Sigma}({\cal L}_{\vec N}A_b^i)\;{\bf E}^b_i 
\end{eqnarray}
where have defined $D^{\bf E}({\vec N})$ as
\begin{equation}
D^{\bf E}({\vec N}):= \int_{\Sigma}({\cal L}_{\vec N}A_b^i)\;(E^b_i- {\bf E}_i^b).
\label{defdnbfe}
\end{equation}
Next, let $\delta$ be some small parameter and let $D^{\bf E}_{\delta}({\vec N})$ 
be an approximant to $D^{\bf E}({\vec N})$ so that 
\begin{equation}
\lim_{\delta\rightarrow 0}D^{\bf E}_{\delta}({\vec N}) = D^{\bf E}({\vec N})
\end{equation}
We define the quantity $D_{\delta}({\vec N})$ as:
\begin{equation}
D_{\delta}({\vec N})= 
\frac{(e^{i \int_{\Sigma} { A}^i_b    \left(\phi ({\vec N}, \delta)^*({\bf E}_i^b)-{\bf E}_i^b \right)}      
(1 +i D^{\bf E}_{\delta}({\vec N}  ) )- 1}{i\delta} ,
\label{dndelta}
\end{equation}
where $\phi ({\vec N}, \delta)$ is the diffeomorphism generated by the shift vector field 
${\vec N}$ so that $\phi ({\vec N}, \delta)$ translates points in $\Sigma$ along the orbits of 
${\vec N}$ by an affine amount $\delta$.
It is then straightforward to verify that 
\begin{equation}
\lim_{\delta\rightarrow 0}D_{\delta}({\vec N}) = D({\vec N})
\end{equation}
so that the expression (\ref{dndelta}) defines an approximant to $D({\vec N})$.

Similar to  Reference \cite{mealok}, our strategy is to identify the parameter $\delta$ 
as characterizing the fineness of a triangulation 
$T_{\delta}$ of $\Sigma$ and to  construct the operator corresponding to $D_{\delta}({\vec N})$ in such a way
that it takes the form
\begin{equation}
{\hat D}_{\delta}({\vec N})= \frac{ {\hat U}(\phi ({\vec N}, \delta))- \bf{1} }{i\delta}
\label{d=u-1ks}
\end{equation}
 on the dense domain ${\cal D}_{KS}$
of the finite linear span of KS spinnets.
Accordingly, given any KS spinnet $|s, {\bf E}\rangle$ we construct 
the quantization of each of the factors 
$(1 +i \delta D^{\bf E}_{\delta}({\vec N}  ))$ 
and 
$e^{i \int_{\Sigma} { A}_b (\phi ({\vec N}, \delta)^*({\bf E}^b)-{\bf E}^b)}$ 
in equation (\ref{dndelta}) in turn, as follows.

%

We note that:\\
\noindent (i)
Equation (\ref{eks}) may be rewritten as:
\begin{equation}
({\hat E}_S(f) -{\bf E}_{S}(f)) |s,{\bf E}_{}\rangle = |{\hat E}_S(f)s,{\bf E}_{}\rangle .
\label{fluxdiff}
\end{equation}
Equation (\ref{fluxdiff}) shows that the action of the operator
${\hat E}_S(f)-{\bf E}_S(f)$ 
on the state $|s, {\bf E}\rangle$ in the KS representation is isomorphic to  the action 
of the electric flux operator ${\hat E}_S(f)$  on the spinnet $|s\rangle$ in the standard LQG representation.\\
\noindent (ii) From equations (\ref{fluxdef}) and (\ref{bfluxdef}), it follows that 
\begin{equation}
{\hat E}_S(f)-{\bf E}_S(f) = \int_Sf^i ({\hat E}^a_i- {\bf E}^a_i)\eta_{abc}.
\label{fluxdiffdef}
\end{equation}

Next, recall that in Reference \cite{mealok} a finite triangulation approximant $D^{LQG}_{\delta}({\vec N})$ 
to the diffeomorphism 
constraint was obtained in terms of  holonomies along  small edges  and fluxes through small
surfaces, the size of these small objects going to zero in the continuum limit. 
It is easy to see that if, in the expression for $D^{LQG}_{\delta}({\vec N})$ in Reference \cite{mealok},  
we substitute each occurrence of an electric flux through a small surface 
by the {\em difference} of the electric flux
and the background electric flux through the same surface, we obtain a finite triangulation approximant to 
$D^{\bf E}({\vec N})$. 
Indeed this can be readily inferred 
from equations (\ref{defdnbfe}), (\ref{fluxdiffdef}) together with  a quick perusal of Reference \cite{mealok}.
\footnote{In \cite{mealok}, the diffeomorphism constraint is split into a vector constraint and Gauss Law contribution.
The finite triangulation vector constraint approximant
is dealt with in rigorous detail and the finite triangulation Gauss Law approximant semi- heuristically, the 
assumption being that the treatment of the latter can be improved so as to construct it in terms of 
holonomy- flux variables. Under this assumption we expect that our treatment here 
should  follow by replacing, in both approximants,
every occurrence of the electric flux by the flux difference.}

Let us denote the phase space dependence of any function $O$ by $O[A,E]$. Using this notation, the discussion
of the previous paragraph implies that we may choose the 
finite triangulation approximant $D^{\bf E}_{\delta}({\vec N})\equiv D^{\bf E}_{\delta}({\vec N})[A,E]$ as:
\begin{equation}
D^{\bf E}_{\delta}({\vec N})[A,E]= D^{LQG}_{\delta}({\vec N})[A,E- {\bf E}].
\end{equation}
Next, recall that in standard LQG we have, from Reference \cite{mealok} that:
\begin{equation}
{\hat D}^{LQG}_{\delta}({\vec N})|s\rangle= \frac{ {\hat U}(\phi ({\vec N}, \delta))- \bf{1} }{i\delta}|s\rangle ,
\label{d=u-1lqg}
\end{equation}
where the finite triangulation operator ${\hat D}^{LQG}_{\delta}({\vec N})$  is defined by replacing the
holonomies and fluxes in the classical approximant, $D^{LQG}_{\delta}({\vec N})[A,E]$,  by the corresponding operators
in standard LQG.

It then follows from (i), (ii) above (see equations (\ref{fluxdiff}), (\ref{fluxdiffdef})) and equation (\ref{d=u-1lqg})
that we have that 
\begin{equation}
{\hat D}^{\bf E}_{\delta}({\vec N})|s,{\bf E}_{}\rangle 
= |\frac{ {\hat U}(\phi ({\vec N}, \delta))- \bf{1} }{i\delta} s, {\bf E}_{}\rangle ,
\label{d=u-1lqgks}
\end{equation}
which implies that the action of the second  factor in equation (\ref{dndelta}) on the KS spinnet $|s,{\bf E}_{}\rangle$
 is:
\begin{equation}
(1 +i\delta  {\hat D}^{\bf E}_{\delta}({\vec N}  ))|s,{\bf E}_{}\rangle
 = |{\hat U}(\phi ({\vec N}, \delta)) s,\; {\bf E}_{}\rangle .
\label{1idbfeks}
\end{equation}

Next, we turn our attention to the first factor, 
$e^{i\int_{\Sigma} \left( \phi ({\vec N}, \delta)^*({\bf E}^b)-{\bf E}^b \right)A_b}$, in equation (\ref{dndelta}).
From equation (\ref{defbe}) this factor is just a background exponential with background field \\
$(\phi ({\vec N}, \delta)^*({\bf E}_i^b)-{\bf E}_i^b)$ so that we have 
\begin{equation}
H_{ \phi ({\vec N}, \delta)^*({\bf E})-  {\bf E} } (A)= 
 e^{i\int_{\Sigma}A_b^i (\phi ({\vec N}, \delta)^*({\bf E}^b_i)-{\bf E}^b_i)} .
\label{hediff}
\end{equation}
It then follows from equations (\ref{hediff}), (\ref{dndelta}), (\ref{defheks}) and (\ref{1idbfeks})
that 
\begin{eqnarray}
{\hat D}_{\delta}({\vec N})|s,{\bf E}_{}\rangle  &=&
\frac{
{\hat H}_{ \phi ({\vec N}, \delta )^*({\bf E})-  {\bf E} }
(1 +i \delta {\hat D}^{\bf E}_{\delta }({\vec N}  ) )- 1}{i\delta} |s,{\bf E}_{}\rangle \\
&=&\frac{
{\hat H}_{ \phi ({\vec N}, \delta)^*({\bf E})-  {\bf E} }
|{\hat U}(\phi ({\vec N}, \delta)) s,\; {\bf E}_{}\rangle  - |s,{\bf E}_{}\rangle }{i\delta}\\
&=& \frac{|{\hat U}(\phi ({\vec N}, \delta)) s,\;  \phi ({\vec N}, \delta)^*({\bf E})  \rangle 
- |s, {\bf E}\rangle}{i\delta}
\\
&=& 
\frac{ {\hat U}(\phi ({\vec N}, \delta))- \bf{1} }{i\delta}|s,{\bf E}_{}\rangle
\end{eqnarray}
where the final line follows from the unitary action of finite diffeomorphisms in the KS representation
\cite{hanno}.
Thus, we have obtained equation 
(\ref{d=u-1ks}) in terms of the holonomy, flux and, {\em crucially} the background exponential
operators of the KS representation. Indeed, as explicitly seen above, it is the background exponential 
contribution which moves the background field by the diffeomorphism 
$\phi ({\vec N}, \delta)$.

The next step would be to take the continuum limit of the above equation on 
an appropriate  generalisation of the  Lewandwoski- Marolf  habitat. We expect that, in analogy to the
standard LM habitat,
elements of this ``LMKS'' habitat would be complex linear mappings on  the finite linear span of 
KS spinnets ${\cal D}_{KS}$. Each habitat state would be a finite linear combination of 
certain elementary basis states $\Psi_{f,g,[(s,{\bf E})]}$.
Here the label $[(s, {\bf E})]$ is the diffeomorphism equivalence class of the KS spinnet labels $(s,{\bf E})$
so that every element of $[(s, {\bf E})]$ is the image of $(s,{\bf E})$ by some diffeomorphism.
The label $f$ 
is the   `vertex smooth' function associated with $s$ (see \cite{donjurek}) and $g$ is a complex functional
(with suitable functional differentiability properties) 
on the space of density weight one Lie algebra valued vector fields. 
Since the main purpose of this work was to demonstrate the key role of the background exponential 
operators in the KS representation, we leave the working out of these ideas for the future.

\section{Concluding Remarks}

In this paper we showed how to construct finite triangulation approximants to the diffeomorphism constraint 
operator in the KS representation. Since the background electric fields labelling the KS spinnets
are mapped to their diffeomorphic images by the unitary action of finite diffeomorphisms and since 
the holonomy- flux operators do not change this background field label, it is imperative to introduce
phase space functions whose operator correspondents {\em do} change these labels, and to use such
operators  in the construction of the desired finite triangulation approximants. Here we showed that these
operators are exactly the background exponential operators constructed in section 2.

It turns out that the background exponentials play a crucial role in the imposition of the Gauss Law and diffeomorphism
constraints in the KS representation. These results \cite{miguelme} constitute an improvement over the pioneering 
works of Koslowski and Sahlmann \cite{kos,hanno,ks} and we shall report them elsewhere.
Clearly the inclusion of  background exponentials constitutes an enlargement of the 
holonomy flux algebra of LQG. Preliminary work \cite{kscstar} suggests that, just as in LQG, 
the commutative part of this  algebra
generated by the holonomies and background exponentials
 can be completed to a commutative $C^*$ algebra, that
the Gel'fand spectrum of this $C^*$ algebra is the topological completion of the space of smooth connections
and that the KS Hilbert space can be obtained as the space of square integrable functions over the spectrum with
respect to an appropriately defined `KS' measure. We will report on these results, subject to confirmation, elsewhere.

Our aim is to generalise the KS representation to the asymptotically flat case using many of the results 
discussed above so as to incorporate the asymptotically flat 
boundary conditions on the phase space variables in quantum theory. In particular, since the triad field 
asymptotes to a smooth, flat triad at infinity, we expect that the smooth background triad fields of the 
KS representation facilitate the imposition of such boundary conditions in quantum theory.
While we do expect progress towards the construction of the KS representation
at the kinematic, $SU(2)$ gauge invariant and spatial diffeomorphism invariant levels in the spatially compact and 
asymptotically
flat cases, 
the construction and imposition of the  Hamiltonian constraint, just as in LQG, remains an open issue.
In order to build intuition for dynamical issues in the case of a compact Cauchy slice, 
Sengupta \cite{sandipanc} has analysed the 
KS representation for the toy model of Parameterised Field Theory (PFT) on the Minkowskian cylinder, 
successfully generalising the 
considerations of Reference \cite{polypft} to the case of the KS representation. On the other hand, PFT on the Minkowskian plane provides a simplified 
setting for a generalization of 
the KS  representation to asymptotically flat gravity. It turns out that 
the boundary conditions on the (spatial derivatives of the) embedding variables in planar PFT 
bear a resemblance to  those for the triad field in asymptotically flat gravity.
Work on this generalization of the KS representation to planar PFT is also in progress by Sengupta \cite{sandipannc}.\\

\noindent {\bf Acknowledgements:} 
I am grateful to Hanno Sahlmann for going through a preliminary sketch of the arguments presented above as
well as for discussions. 
It is a pleasure to thank Miguel Campiglia and 
Sandipan Sengupta for useful discussions and Alok Laddha for going through a draft version of this work as well as for
useful discussions.


\begin{thebibliography}{999}
\bibitem{discrete}
C. Rovelli and L. Smolin, {\sl Nucl.Phys.}{\bf B442} (1995) 593-622, Erratum-ibid. {\bf B456}, 753 (1995);
A. Ashtekar and J. Lewandowski, 
{\sl Class.Quant.Grav.}{\bf 14}  A55 (1997), ibid {\sl Adv.Theor.Math.Phys.}{\bf 1},  388 (1998).









\bibitem{kos}	
`Dynamical Quantum Geometry', T. Koslowski,  e-Print: arXiv:0709.3465 .



\bibitem{hanno} H. Sahlmann, 
{\sl Class.Quant.Grav.}{\bf  27}, 225007 (2010).



\bibitem{mealok} A. Laddha and M. Varadarajan, 
{\sl Class.Quant.Grav.}{\bf 28}, 195010 (2011).



\bibitem{miguelme} M. Campiglia and M. Varadarajan, In preparation.

\bibitem{sandipanc} S. Sengupta, eprint: arXiv:1306.6013 [gr-qc].



\bibitem{donjurek} J. Lewandowski and D. Marolf, {\sl Int.J.Mod.Phys.}{\bf D7}, 299 (1998). 



\bibitem{ks} T. Koslowski and H. Sahlmann, {\sl SIGMA}{\bf 8}, 026 (2012).


\bibitem{kscstar} M. Campiglia, A. Laddha and M. Varadarajan, In progress.




\bibitem{sandipannc} S. Sengupta, In progress.




\bibitem{polypft}  A. Laddha and M. Varadarajan, {\sl Class.Quant.Grav.}{\bf  27}, 175010 (2010).


\end{thebibliography}
\end{document}